\documentclass[draft]{agujournal2019}
\usepackage{url} %this package should fix any errors with URLs in refs.
\usepackage{lineno}
\usepackage[inline]{trackchanges} %for better track changes. finalnew option will compile document with changes incorporated.
\usepackage{soul}
\usepackage{amsmath,amssymb}
\usepackage{pdflscape} 
\usepackage{array} % required for text wrapping in tables

\draftfalse

\journalname{Journal of Geophysical Research: Space Physics}

\begin{document}

\title{Debye-scale electrostatic waves across quasi-perpendicular shocks}

\authors{Ahmad Lalti\affil{1,2,3}, Yuri V. Khotyaintsev\affil{2,3}, Daniel B. Graham\affil{2}, Andris Vaivads\affil{4,5}}

\affiliation{1}{Northumbria University, Newcastle upon Tyne, UK}
\affiliation{2}{Swedish Institute of Space Physics, Uppsala, Sweden}
\affiliation{3}{Department of Physics and Astronomy, Uppsala University, Uppsala, Sweden}
\affiliation{4}{Division of Space and Plasma Physics, KTH Royal Institute of Technology, Stockholm, Sweden}
\affiliation{5}{Ventspils University of Applied Sciences, Ventspils, Latvia}

\correspondingauthor{Ahmad Lalti}{ahmad.lalti@northumbria.ac.uk}

\begin{keypoints}
\item We investigate the evolution of electrostatic wave properties across four quasi-perpendicular shocks using multi-probe interferometry.
\item For all four shocks most analyzed waves occur at Debye scales, have a plasma frame frequency $<f_{pi}$, and phase speed $<c_s$.
\item The waves are highly oblique to $\mathbf{B}$ in the shock transition region and are predominantly field-aligned outside of it.
\end{keypoints}

\begin{abstract}
The evolution of the properties of short-scale electrostatic waves across collisionless shocks remains an open question. We use a method based on the interferometry of the electric field measured aboard the magnetospheric multiscale spacecraft to analyze the evolution of the properties of electrostatic waves across four quasi-perpendicular shocks, with $1.4 \leq M_A \leq 4.2$ and $66^\circ \leq \theta_{Bn} \leq 87^\circ$. Most of the analyzed wave bursts across all four shocks have a frequency in the plasma frame $f_{pl}$ lower than the ion plasma frequency $f_{pi}$ and a wavelength on the order of 20 Debye lengths $\lambda_D$. Their direction of propagation is predominantly field-aligned upstream and downstream of the bow shock, while it is highly oblique within the shock transition region, which might indicate a shift in their generation mechanism. 
The similarity in wave properties between the analyzed shocks, despite their different shock parameters, indicates the fundamental nature of electrostatic waves for the dynamics of collisionless shocks.
\end{abstract}

\section{Introduction}

Debye scale electrostatic waves play an integral role in the dynamics of collisionless shocks. They are believed to provide the anomalous dissipation necessary to maintain the shock \cite{sagdeev1966cooperative,galeev1976collisionless,papadopoulos1985microinstabilities,wilson2014quantified}. At Earth's bow shock, their electric fields can reach 100s of mV/m, which allows them to efficiently interact and scatter electrons \cite{balikhin1998study,bale2007measurement,see2013non,kamaletdinov2022quantifying,kamaletdinov2024nonlinear} playing a major role in electron thermalization across the shock. The microphysics of collisionless shocks is dependent on the macroscopic shock paramteres \cite{balogh2013physics}. For example, supercritical shocks reflect ions to maintain the shock, while subcritical shocks rely mainly on anomalous dissipation. To understand the role that Debye scale electrostatic waves play in the microphysics of collisionless shocks, there is a need to identify the wave modes excited along with their properties, their occurrence rate, and their dependence on the macroscopic shock parameters.

Debye scale electrostatic waves have been observed since the early days of in situ observations of collisionless shocks \cite{rodriguez1975electrostatic}. However, because of their small spatial size and fast temporal evolution, the identification and characterization of the various wave modes using in situ observations have been challenging.
Nevertheless, different techniques have been employed to characterize various wave modes, such as ion acoustic waves \cite{fuselier1984short,balikhin2005ion,hull2006large,goodrich2018mms,vasko2022ion}, bipolar structures as a signature of electron and ion phase space holes \cite{bale2002electrostatic,vasko2020nature,wang2020electrostatic}, and electron cyclotron drift instability (ECDI) waves \cite{wilson2010large,breneman2013stereo,wilson2014quantified,cohen2020rapid}.

The magnetospheric multiscale (MMS) spacecraft \cite{Burch2016} is arguably one of the most advanced spacecraft to explore Earth's bow shock. It uses the double probe technique \cite{pedersen1998electric} to measure the 3D electric field with two spin plane double probes (SDP) \cite{lindqvist2016spin} and an axial double probe (ADP) \cite{ergun2016axial} with probe-to-probe separations of 120m and 30m respectively constituting the electric field double probe (EDP) instrument. It has been shown that the electric field measurement by EDP of waves with a wavelength on the order of the probe-to-probe separation is systematically biased towards the spacecraft's axial direction, which will affect wave mode determination \cite{vasko2018solitary,goodrich2018mms,steinvall2022applicability,lalti2023short}. Because of this technical limitation, identifying which are the dominant electrostatic wave modes and their properties remains open.

\citeA{lalti2023short} developed a method to reliably measure the 3D dispersion relation of short-scale electrostatic waves, based on spin-plane interferometry of the electric field, and benchmarked their method on synthetic data and ion-acoustic waves in the solar wind.
Here, we apply this method to study the evolution of the properties of short-scale electrostatic waves excited across four quasi-perpendicular shocks with different Mach numbers $M_A$ and shock geometry $\theta_{Bn}$.

\section{Dataset}

 The macroscopic shock parameters of the chosen shocks are listed in Table \ref{tab:tab1}. To show if the characteristics of electrostatic waves across quasi-perpendicular shocks vary with shock properties, we have chosen the analyzed shocks such that two shocks are supercritical (shocks 1 and 2), and two are subcritical (shocks 3 and 4). And in each category one has a perpendicular geometry while the other is more oblique, as can be seen in Table \ref{tab:tab1}. Shocks 1 and 2 were chosen from the database of \citeA{lalti2022database}, while shocks 3 and 4 were analyzed in \citeA{graham2024ion,graham2024structure}.

\begin{table}
    \centering
    \begin{tabular}{|c|>{\centering\arraybackslash}p{0.15\linewidth}|c|c|>{\centering\arraybackslash}p{0.1\linewidth}|>{\centering\arraybackslash}p{0.15\linewidth}|>{\centering\arraybackslash}p{0.1\linewidth}|>{\centering\arraybackslash}p{0.1\linewidth}|l|}
    \hline
        Shock $\#$ & Time of crossing & $M_A$ & $\theta_{Bn}$ & Spacecraft used & $\#$ of subintervals with $|E|>E_{thresh}$ &$E_{thresh}$
$(mV/m)$& $\#$ of analyzed wavebursts &$\frac{Column \;8}{Column \; 6}\times 100$\\
    \hline    
         1 & 2017-11-24 23:20:15.44 & $4.2$ & $82^\circ$ & MMS1 and MMS3 & 3420&1& 803  &23.5\%\\
    \hline     
         2 & 2017-11-02 08:29:17.17 & $4.2$ & $66^\circ$ & MMS1 & 957&1& 431  &45.03\%\\
    \hline
         3 & 2023-04-24 03:50:12.25 & $1.4$ & $87^\circ$ & MMS1 & 675&1& 171  &25.33\%\\
    \hline
         4 & 2023-04-24 04:15:24.98 & $1.6$ & $70^\circ$ & MMS1 & 807&5& 212  &26.27\%\\
    \hline
    \end{tabular}
    \caption{Properties of the four shocks analyzed.}
    \label{tab:tab1}
\end{table}

\section{Method}
\label{method}

To study the evolution of short-scale electrostatic waves across the selected shocks we use the method developed by \citeA{lalti2023short}. Using the probe-to-spacecraft potential data product provided by the EDP instrument at a sampling frequency of $\sim 8$ kilo samples per second (kS/s), \citeA{lalti2023short} calculated two electric field components in the spin plane of the spacecraft, each at two different locations in space. By applying a wavelet transform and calculating the phase shift between the two spatial measurements of the same component, they obtained a frequency-wavenumber (f-k) spectrogram. From this spectrogram, they determined the spin plane components of the dispersion relation of the analyzed wave. By modeling the response of the 3 different double probes on MMS to a short-wavelength plane wave at different propagation directions, they were able to obtain the frequency-dependent 3D wave vector (both magnitude and direction of propagation) of the analyzed wave.

For the method to work on any particular wave burst two conditions must be met. The first is that the electric field should have a large enough component in the spin plane for the probes to measure an electric field signal and the second is that the wave should have a coherent and dispersive signal in the f-k spectrogram. In other words, if the wave vector is colinear with the axial direction of the spacecraft, or if the wave is a superposition of multiple interacting wave modes of similar amplitude then the method will fail. Finally, we mention that the method can resolve wavelengths in the range $[50,\; 500]$ m, or $\sim [5, \; 50] \lambda_D$ in the solar wind. For more details, the reader is referred to \citeA{lalti2023short}.

%Figure 1: top panels E and B fields, zoom in on 1 waveburst, two panels for dispersion relation with fit.
\begin{figure}[htbp]
    \centering
    \includegraphics[width=\linewidth]{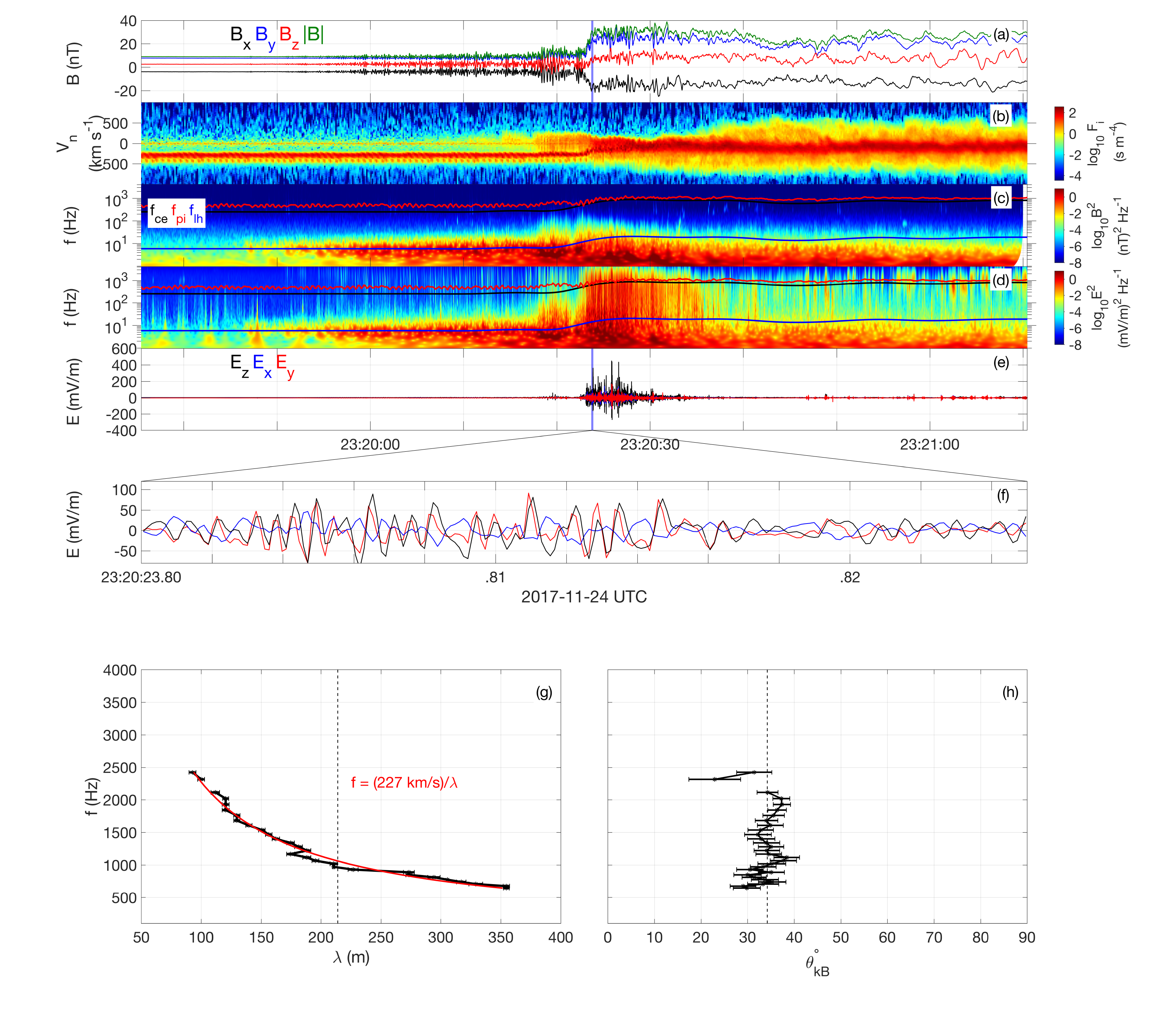}
    \caption{Overview of the MMS shock crossing on November 24, 2017. Panel (a) shows the magnetic field in the GSE coordinate system. Panel (b) shows the reduced ion velocity distribution function in the normal direction. Panels (c-d) show the power spectral density of the magnetic and electric fields. Panel (e) shows the electric field in the GSE coordinate system. Panel (f) is a zoom-in on one wave burst in the shock ramp. Panels (g-h) show the f versus $\lambda$ and f versus $\theta_{kB}$ in black respectively. The dashed vertical lines represent the weighted mean $\lambda$ and $\theta_{kB}$ representative for this wave burst. The red curve in panel (g) is the least square fit of the data to $f = V_{ph}/\lambda$. }
    \label{fig:fig1}
\end{figure}

As an example application for the method, we show in Figure \ref{fig:fig1} an overview of the shock crossing by MMS on November 24, 2017 (shock $1$). MMS was initially in the solar wind, it crossed the shock at around 23:20:45 UT and then moved into the magnetosheath. This is a supercritical quasi-perpendicular shock with $M_A = 4.2$ and $\theta_{Bn} = 82^\circ$. In panel (b) we plot the reduced ion distribution along the normal direction. It shows the typical signature of a supercritical shock, with a cold solar wind beam upstream of the shock, a reflected ion component with $V_n>0$ on top of the solar wind component in the foot, and the thermalized downstream. The electric field amplitude at the shock reaches values as high as 500 mV/m peak-to-peak. The PSDs in panels (c) and (d) show the signature of upstream precursor electromagnetic waves around the lower hybrid frequency, which were shown by \citeA{lalti2022whistler} to be obliquely propagating whistler waves with a wavelength around the ion inertial length. In addition, by comparing the magnetic and electric field PSDs (panels (c) and (d) respectively), the blue shading above $\sim 100$ Hz indicates that there is no power in the magnetic field, and hence the oscillations are predominantly electrostatic.

Panel (f) is a zoom-in on a wave burst within the ramp with an amplitude reaching 100 mV/m peak-to-peak. Panels (g-h) are the dispersion relation of that wave burst obtained by applying the method of \citeA{lalti2023short}. In Panel (g) we show in black the measured frequency versus wavelength curve, while in red is the best fit to the acoustic-like dispersion relation $f = V_{ph}/\lambda$. There is good agreement between the measured and best-fit dispersion relation with a best-fit phase speed in the spacecraft frame of $= 227 \pm 30$ km/s. Panel (h) shows a plot of the frequency versus $\theta_{kB}$. In both panels, the vertical dashed line is the weighted mean wavelength and $\theta_{kB}$ with the weights taken as the value of the power spectral density. To calculate the plasma frame $f_{pl}$ frequency we use
\begin{equation}
    f_{pl} = f_{sc} - \frac{\mathbf{k}\cdot\mathbf{V}}{2\pi},
    \label{doppler}
\end{equation}
with $\mathbf{k}$ being the wave vector, and $\mathbf{V}$ being the local bulk velocity. This wave has an average frequency in the plasma frame normalized to the ion plasma frequency $f_{pl}/f_{pi} = 0.1 \pm 0.07$, an average wavelength normalized to the Debye length $\lambda/\lambda_{D} = 19 \pm 7$, and an average $\theta_{kB} = 35 \pm 3^\circ$. 
The Debye length $\lambda_D$ for this wave burst is $\sim 11$ m. In addition, its average plasma-frame phase speed $V_{pl} = V_{sc} - \frac{\mathbf{k}\cdot\mathbf{V}}{|\mathbf{k}|} = -13 \pm 12$ km/s, with $V_{sc}$ is the spacecraft frame phase speed. The obtained value for $V_{pl}$ indicates that this wave is almost phase-standing in that frame of reference. The acoustic-like dispersion relation seen in Panel (g) along with $V_{pl}$ being near zero reflects the fact that the Doppler shift term in equation \ref{doppler} dominates the observed dispersion relation and this wave is mainly convected with the plasma. 

To analyze the evolution of the properties of the different electrostatic waves observed across the four shocks, we divide the time interval of each shock crossing into 25 ms long subintervals, corresponding to a minimum frequency of 40 Hz. This is chosen so that waves with frequency $>100$ Hz are well captured. For each subinterval, we plot the f-k spectrogram and select the ones that show a coherent signal in the f-k spectrogram of the spin-plane electric field components. Finally, for each selected waveburst, we calculate the various wave properties using the same approach as that for the waveburst shown in Figure \ref{fig:fig1}. 

\section{Results}
\label{results}
Using the method described above we analyze the evolution of electrostatic wave properties across the four shocks listed in Table \ref{tab:tab1}. For Shock 1, we analyze and compare the observations of MMS1 and MMS3. We find consistent wave properties between the two, in addition, given that the separation between the two spacecraft was $\sim 2 \times 10^4$ m, and the wavelength of the analyzed waves is of the order of $\sim 10^2$ m, we analyze the two datasets in unison to increase our statistics. For Shocks 2-4, we only use observations from MMS1. In the sixth column of Table \ref{tab:tab1}, we show the number of subintervals that we consider for analyses filtered in such a way that the high-pass filtered electric field magnitude is larger than a threshold value $E_{thresh}$ shown in column 7 of Table \ref{tab:tab1}. This threshold electric field value is to ensure that in the time interval considered we have an observable electric field disturbance. Note that for shock 4 it takes more time to cross from the solar wind to the magnetosheath, therefore, to make the manual labor of the wave analysis process manageable without affecting the statistics for the shock we chose a higher value for $E_{thresh}$. The fraction of the analyzed intervals compared to all intervals showing wave activity is shown in column 8 with most of the shocks having a value $\sim 25\%$. The remaining non-analyzed subintervals show nonlinear wave activity whose f-k spectrogram shows no coherent signature, and therefore, are too complex for the method to analyze.
Note that for all analyzed wavebursts for all four shocks, we do not differentiate between waveforms, bipolar, and monopolar structures.

\subsection{Shock 1}

\begin{figure}[htbp]
    \centering
    \includegraphics[width=1\linewidth]{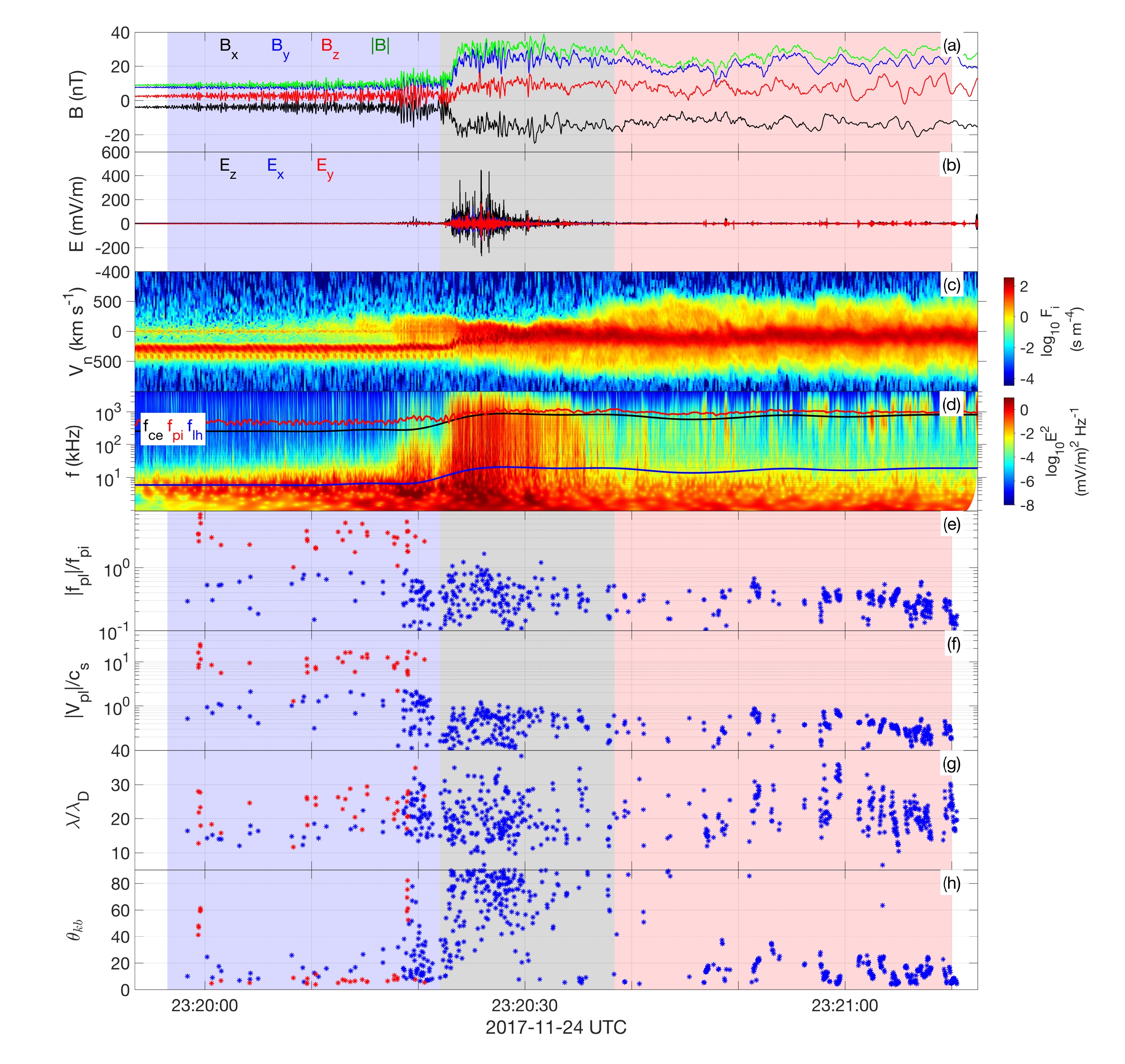}
    \caption{Results of the analysis of 803 wave bursts from MMS1 and MMS3 across the shock. Panels (a-d) are the magnetic and electric field, the ion VDF in the normal direction, and the electric field PSD respectively. Panel (e) shows the absolute value of the frequency in the plasma frame normalized by the ion plasma frequency. Panel (f) shows the absolute value of the phase speed in the plasma frame normalized by the local sound speed. Panels (g-h) show the wavelength and $\theta_{kB}$. The high-frequency population of waves has been highlighted in red in the upstream. The upstream, shock transition region and the downstream have been highlighted in blue, black, and red respectively.}
    \label{fig:fig2}
\end{figure}
 The first shock analyzed is that shown in Figure \ref{fig:fig1}. Using the method described above we were able to determine the wave properties of 803 wavebursts across the shock out of 3420 subinterval considered.
Figure \ref{fig:fig2} shows the result of this analysis. We plot the magnetic and electric fields time series, the ion VDF reduced along the normal direction, and the electric field PSD in panels (a-d) respectively. In panels (e-h) we plot $|f_{pl}|/f_{pi}$ and $|V_{pl}|/c_s$ the absolute value of the frequency normalized to the local ion plasma frequency and the phase speed normalized to the local sound speed both in the plasma rest frame, $\theta_{kB}$ the angle between the wave vector and the background magnetic field, and $\lambda/\lambda_{D}$ the normalized wavelength for each of the 803 wave bursts analyzed. The sound speed $c_s$ is equal to
\begin{equation}
c_s = \frac{T_e+3T_i}{m_p},    
\label{cs}
\end{equation}
with $T_e$ and $T_i$ being the electron and ion temperatures and $m_p$ the proton mass. Since the Fast Plasma Investigation (FPI) Dual Ion Spectrometer (DIS) aboard MMS \cite{pollock2016fast} overestimates the ion temperature in the solar wind, to calculate $c_s$ we use the ion temperature from OMNI for the upstream (blue-shaded) interval $T_{iu} \sim 3$ eV, and the MMS measurement in the shock transition region (STR) and the downstream (black and red shaded) intervals. This will result in an accurate sound speed for the upstream and the downstream. The STR is a mixture of incoming solar wind and shock-reflected ions. Because of the complex distribution function in that region, Eq. \ref{cs} is only an approximation to the actual sound speed.

We find that the plasma frame frequency $\left(f_{pl}\right)$ for most of the wave bursts analyzed $\left(\sim 90\% \right)$ is below the local ion plasma frequency $\left(f_{pi}\right)$(see Figure \ref{fig:fig2}e). The remaining $10\%$ are almost entirely in the upstream interval (blue-shaded). In that interval, we observe two populations of waves, the first (plotted in red) is characterized by a higher frequency $f_{pl}/f_{pi}>1$ with median values of the wave properties $med\left(f_{pl}/f_{pi}\right) = 3 \pm 1$, $med\left(V_{pl}/c_s\right)=10 \pm 5$, $med\left(\lambda/\lambda_D\right) = 27 \pm 6$ and $med\left(\theta_{kB}\right) = 8^\circ \pm 20^\circ$ which we will explore further below. The second population (plotted in blue) is characterized by a lower frequency $f_{pl}/f_{pi}<1$, with median wave properties  $med\left(f_{pl}/f_{pi}\right) = 0.3 \pm 0.2$, $med\left(V_{pl}/c_s\right)=1 \pm 0.6$, $med\left(\lambda/\lambda_D\right) = 20 \pm 4$ and $med\left(\theta_{kB}\right) = 16^\circ \pm 15^\circ$. Both populations propagate within $\sim 20^\circ$ from the background magnetic field, however, the lower frequency waves tend to be more oblique.

Moving into the STR (black shaded interval) the higher frequency population is not observed anymore, and the remaining wave bursts have median properties $med\left(f_{pl}/f_{pi}\right) = 0.3 \pm 0.3$, $med\left(V_{pl}/c_s\right)=0.3 \pm 0.3$, $med\left(\lambda/\lambda_D\right) = 20 \pm 6$ and $med\left(\theta_{kB}\right) = 72^\circ \pm 24^\circ$. 
Finally, in the downstream (red shaded interval) the lower frequency waves persist with median properties $med\left(f_{pl}/f_{pi}\right) = 0.3 \pm 0.1$, $med\left(V_{pl}/c_s\right)=0.3 \pm 0.2$, $med\left(\lambda/\lambda_D\right) = 21.5 \pm 5$ and $med\left(\theta_{kB}\right) = 13^\circ \pm 10^\circ$. The lower frequency waves have comparable properties across the shock except for one main difference, which is $\theta_{kB}$. Outside the STR the waves are much more aligned with the background magnetic field compared to within the STR where $\theta_{kB}$ becomes almost perpendicular. This change in the wave properties across the shock suggests a change in the wave generation mechanism as we will discuss in the next section.

\subsection{Shocks 2-4}

\begin{landscape}
\begin{figure}[htbp]
    \centering
    \includegraphics[width=1.1\linewidth, height=1.1\textheight, keepaspectratio]{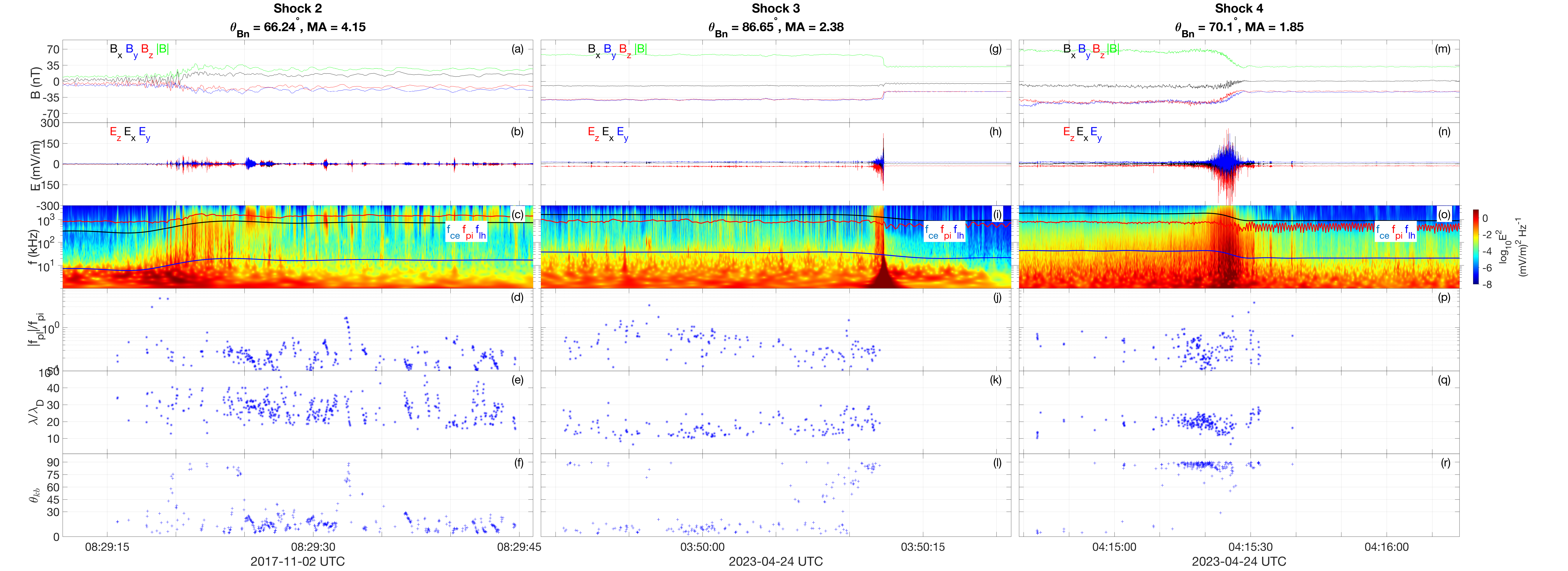}
    \caption{Results of the analysis of the electrostatic waves across shocks 2,3 and 4. Panels (a-f) show respectively, the magnetic and electric fields, the electric fields PSD, and the wave properties, $|f_{pl}|/f_{pi}$, $\lambda/\lambda_D$, and $\theta_{kB}$ for shock 2. While Panels (g-l) and (m-r) show the same quantities for shocks 3 and 4. }
    \label{fig:fig4}
\end{figure}
\end{landscape}

The results for the analysis of the electrostatic waves across the remaining shocks are shown in Figure \ref{fig:fig4}. Panels (a-f) show respectively, the magnetic and electric fields, the electric fields PSD, and the wave properties, $|f_{pl}|/f_{pi}$, $\lambda/\lambda_D$, and $\theta_{kB}$ for shock 2. While Panels (g-l) and (m-r) show the same quantities for shocks 3 and 4.

As mentioned above, each of the four shocks is chosen to have a different combination of the $\left(M_A, \theta_{Bn}\right)$ tuple. Despite the differences in parameters, the properties of the electrostatic waves across the four shocks are very similar. The high-frequency waves observed in shock 1 are mostly absent for shocks 3 and 4 where the upstream waves had too small of amplitude to be analyzed by our method. In panel (d), there are 3 points with $f_{pl}>f_{pi}$, $\theta_{kB} \sim 0^\circ$, and $\lambda \sim 30 \lambda_D$, which are co-located with electromagnetic whistler waves with frequency around the lower hybrid frequency, consistent with the wave properties of the high-frequency waves of shock 1.
%However, HMFE field measurements for those waves are unavailable; hence, we cannot verify that they exhibit the same substructuring as the waves observed in shock 1.

As for the STR and the downstream, the electrostatic waves have $|f_{pl}|/f_{pi} < 1$, $\lambda/\lambda_D$ ranging from 10 to 40, and $\theta_{kB} \sim 90^\circ$ in the STR and $\sim 0^\circ$ in the downstream. The implication of the similarity of the wave properties between the four shocks despite the differences in shock parameters will be discussed in the next section.

\section{Discussion}
\label{discussion}

We have analyzed the evolution of the properties of Debye-scale electrostatic waves across four quasi-perpendicular shocks. At such scale lengths, not many wave modes exist. We have found that the dominant population of waves for the four shocks analyzed is ion-acoustic wave-like, with $|f_{pl}| \lesssim f_{pi}$, $V_{pl}\sim c_s$ and a $\theta_{kB} \sim 90^\circ$ within the STR and $\sim 0^\circ$ outside of it. 
For each chosen combination of the shock parameters, $\theta_{Bn}$ either perpendicular or oblique and $M_A$ super or sub-critical, the physics underlying the shock is expected to be different \cite{kennel1985quarter,balogh2013physics}. Contrary to subcritical shocks, supercritical shocks are characterized by ion reflection and an upstream foot region that decelerates the solar wind before reaching the main shock. On the other hand, the shock width is highly affected by $\theta_{Bn}$, with oblique shocks being broader than their perpendicular counterpart. Despite all of these differences, the observed properties of the electrostatic waves are similar, indicating the fundamental nature of those electrostatic waves for the dynamics of shocks. This result is essential for future studies looking at the interaction between electrons and ions with the broad spectrum of electrostatic waves excited across quasi-perpendicular shocks.

%%add a sentence saying there are multiple mechanisms that can generate the waves with the properties that we observe talk about the differences in the prediction in each method 
Multiple mechanisms can generate waves with the properties that we observe, each will be active depending on the location with respect to the shock (upstream, downstream, or within the STR), and depending on the values of the macroscopic shock parameters (supercritical versus subcritical and oblique versus perpendicular shocks).
In the ramp, the ions and electrons behave differently due to the differences in their dynamical scale lengths in comparison to the shock width \cite{goodrich1984adiabatic,krauss1989electron}. This difference in dynamics will set up cross-field currents within the ramp that have been shown to provide the free energy necessary to excite obliquely propagating ion acoustic waves \cite{gary1970longitudinalI,gary1970longitudinalII}. In addition, ion-ion drifts across the shock can also result in highly oblique acoustic-like electrostatic waves. In particular, the free energy for such waves can be provided by a proton-alpha drift across the shock \cite{graham2025ionacousticwavesprotonalphastreaming}, or a proton-proton drift between reflected and incoming protons of supercritical shocks \cite{formisano1982ion, akimoto1985ion}. To determine which mechanism dominates for each analyzed shock, one can examine the direction of propagation of the waves with respect to the direction of the drift that provides their free energy. In addition, using the obtained wave-vector and the electric field measurement we can calculate the electric potential of the observed waves. A positive or negative electric potential would imply the formation of an electron or ion hole respectively, which can help constrain the possible mechanisms generating the observed waves. This investigation will be addressed in a future study.

The observed ion acoustic wave properties outside of the STR are consistent with previous observations both in the upstream solar wind \cite{1978Gurnettionacoustic,pivsa2021first,lalti2023short} and the downstream magnetosheath \cite{gallagher1985short,zhu2019composition}. In these regions, ion-acoustic waves can be driven by the turbulent energy cascade, where larger scale waves such as whistler waves can locally modify the distribution function in a way that is favorable for the excitation of the waves \cite{valentini2014nonlinear,saito2017generation,karbashewski2022cascading}. Additionally, in the upstream solar wind, various non-Maxwellian features of the ion and electron distribution functions, such as proton beams and electron strahl, can provide the necessary currents or heat flux to excite ion acoustic waves \cite[and references therin]{verscharen2019multi,verscharen2022electron,boldu2024ion}. Moreover, in the downstream magnetosheath ring ion distributions formed by the shock-reflected ions after they are transmitted to the magnetosheath can also drive ion-acoustic waves \cite{wu1984microinstabilities}.

\begin{figure}[htbp]
    \centering
    \includegraphics[width=0.9\linewidth]{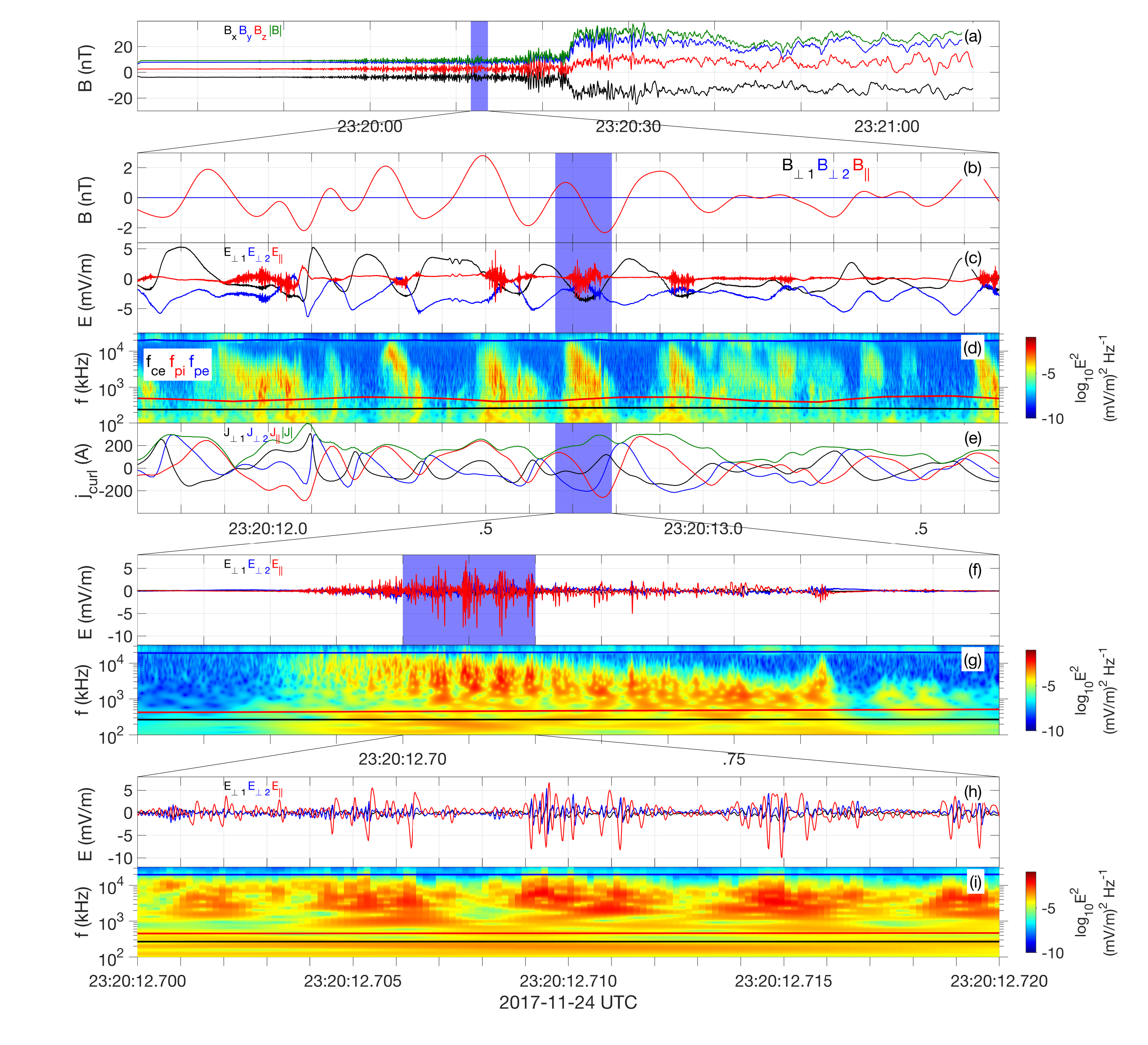}
    \caption{Analysis of the high-frequency wave population. Panel (a) shows the magnetic field in GSE coordinates. Panels (b-e) are a zoom-in on a subinterval showing a few whistler wave periods. They show respectively the magnetic and electric fields in a local field-aligned coordinate system, the HMFE electric field PSD, and the current density in field-aligned coordinates. Panels (f-g) show the electric field and its PSD zoomed in on one of the wave bursts. Panels (h-i) are the same as (f-g) but zoomed in to show the substructuring of the wave burst.}
    \label{fig:fig3}
\end{figure}

For shock 1 and possibly shock 2, we have observed a higher frequency wave population with a frequency larger than the ion plasma frequency and a phase speed larger than the sound speed while having an acoustic-like dispersion relation ($f = V_{ph}/\lambda$). To further analyze those waves we use the HMFE electric field data product from the EDP instrument, provided at a sampling rate of $\sim 65$ kS/s. Figure \ref{fig:fig3} panel (a) shows the magnetic field for the full crossing of shock 1. Zooming in on a time interval near the foot of the shock, panels (b-c) show the magnetic and electric fields respectively rotated to a field-aligned coordinate (FAC) system $\left(\hat{e}_{\perp_1}, \hat{e}_{\perp_1}, \hat{e}_{\parallel}\right)$ based on the local magnetic field $\mathbf{B}$, with $\hat{e}_{\parallel} = \frac{\mathbf{B}}{|\mathbf{B}|}$, $\hat{e}_{\perp_2} = \hat{e}_{\parallel} \times \hat{k}_{wh}$, and $\hat{e}_{\perp_1} = \hat{e}_{\perp_2}\times \hat{e}_{\parallel}$, where $\hat{k}_{wh}$ is the unit vector in the direction of the wave vector of the lower hybrid freequency whistlers determined using the four spacecrat interferometry technique \cite{lalti2022whistler}. The lower frequency oscillations of the fields in panels (b-c) correspond to those whistler waves. Embedded in the fields of the whistlers are the high-frequency electrostatic waves of interest. Panel (d) shows the PSD of the HMFE electric field which shows the full highly dispersive spectrum of the high-frequency waves ranging from $\sim 20$ kHz and going down to slightly above $f_{pi}$. It is worth noting that there is a clear power enhancement near the electron plasma frequency, which is a typical signature of Langmuir waves.

%We calculate the parallel electrostatic potential of the whistler waves
%\begin{equation}
%    \Delta \Phi_{\parallel} = - \int E_{\parallel}V_{ph\parallel} dt,
%\end{equation}
%where $E_{\parallel}$ and $V_{ph\parallel}$ are the wave electric field measured by the EDP instrument on MMS \cite{ergun2016axial,lindqvist2016spin}, and its phase velocity determined using 4 spacecraft interferometry \cite{lalti2022whistler}, both projected on the direction parallel to the background magnetic field. We plot $\Delta \Phi_{\parallel}$ in panel (e). The wave bursts start at local minima in the electrostatic potential (or maxima in the electron potential energy) and their power maximizes at the inflection point on the positive slope phase of the oscillation. 

Furthermore, panel (e) shows the current obtained using the curlometer technique \cite{robert1998accuracy} and rotated to the same field-aligned coordinate system used in panels (b-c). Note that the current obtained using the curlometer technique is measured at the centroid of the tetrahedron. We use the wave vector (in FAC system) of the whistlers determined using multispacecraft interferometry $\mathbf{k}_{wh} =\frac{2\pi}{\lambda_{wh}} \left[0.66,\; 0,\; 0.75\right]$, with $\lambda_{wh} = 70$ km, the wavelength of the whistlers \cite{lalti2022whistler}, to calculate the phase shift necessary to shift the measurement to the location of MMS1. The wave bursts start at local maxima in the parallel current and their power maximizes near the inflection point on the negative slope phase of the oscillation. 
Zooming further in on one of the wave bursts shows additional substructuring at smaller scales (panels f-g) and the occurrence of bipolar structures embedded within the waveform (panels h-i). 

Such properties indicate that they might be linear or nonlinear electron acoustic waves (EAWs) \cite{berthomier2000electron}. Using a 2D PIC simulation of the foot of a quasi-perpendicular shock \citeA{matsukiyo2006microinstabilities} found that a two-step instability can generate EAWs leading to electron thermalization. First, the drift between the reflected ions and the solar wind electrons generates oblique whistler waves at the lower hybrid frequency. Then, when their amplitude is high enough, those whistler waves will trap some of the thermal electrons, forming a locally double peaked distribution function that provides the free energy to excite EAWs. This is consistent with our observations in Figure \ref{fig:fig3} (b-e). In particular, the modulation of the wave occurrence with the phase of maximum amplitude of the parallel current indicates that this current, which is carried by the electrons responding to the parallel electric field of the lower-hybrid frequency whistler waves, likely provides the free energy for the excitation of the higher frequency electrostatic waves.

Furthermore, using a 1D PIC simulation of the interaction between chorus waves in the radiation belt and electrons, \citeA{an2019unified} showed that depending on which population of electrons is trapped, a different wave mode can be excited. They found that when the thermal population is trapped, electron phase space holes will be formed, which leads to the formation of bipolar structures. On the other hand, when the tail of the electron distribution is trapped, this leads to a bump on tail instability and the subsequent generation of Langmuir waves. Finally, if electrons at intermediate energies are trapped, EAWs will be excited through a beam-mode instability. In turn, when the EAWs saturate they disrupt the trapped electrons and form monopolar electric field structures. Examining the substructure of the observed waveforms in panels (f-i) we find that both bipolar electric fields and continuous waveforms are embedded within the full wave burst, consistent with the results of the simulation of \citeA{an2019unified}.

Another generation mechanism that can explain the observed wave properties is that of beam-mode waves discussed by \citeA{onsager1990measurement}, where the field-aligned reflected electrons in the electron foreshock provide the free energy to excite electron-acoustic (equivalently, beam-mode) waves. 
Further evidence regarding the generation mechanism of such waves requires examining the electron distribution function at thermal and suprathermal energies. The latter is resolved by MMS, however, despite providing unprecedented temporal resolution for the electron distributions, the FPI Dual Electron Spectrometer (DES) aboard MMS does not resolve thermal electrons in the solar wind \cite{pollock2016fast}. Future space missions to study the electron scale physics at collisionless shocks will require a higher energy resolution covering thermal electron energies.

We conclude this section by noting that our results are limited by the method employed. In particular, determination of the dispersion relation was restricted to waves with $\lambda \in \left[50\; 500\right]$ m $\sim [5 \; 50] \lambda_D$,  and a spacecraft frequency $f_{sc}<4$ kHz. Wave modes with higher frequencies, wavelengths outside of the resolved range, or waves without a coherent f-k spectrum are not possible to be identified. Those constitute around $\sim 75\%$ of the electrostatic wave activity around the shock that is yet to be explored. Those limitations are to a large extent the result of the EDP instrument design. Further advancement in our knowledge of the electrostatic turbulence at collisionless shocks is contingent on electric field instrument design that can properly resolve Debye scale structures, such as those proposed for ESA's Plasma Observatory \cite{retino2021particle} or NASA's MAKOS \cite{goodrich2023multi}.

\section{Conclusions}
\label{conclusions}

We have used a method based on the interferometry of the spin-plane electric field measured aboard MMS to properly characterize short-scale electrostatic waves and the evolution of their properties across four quasi-perpendicular shock crossings. We find that:
\begin{itemize}
    \item The dominant wave mode excited in the shock environment is characterized by $f_{pl}<f_{pi}$, $V_{pl} \sim c_s$, $\lambda \sim 20 \lambda_D$, and an acoustic-like dispersion relation, consistent with ion-acoustic-like waves.
    \item Within the STR, $\theta_{kB}$ becomes predominantly oblique, indicating the importance of cross-field currents within the ramp for the generation of ion-acoustic-like waves and the dissipation of energy across the shock.
    \item Another wave mode has been identified at shock 1 with  $f_{pl}>f_{pi}$, $V_{pl} > c_s$, $\lambda \sim 24 \lambda_D$, and an acoustic-like dispersion relation, consistent with electron-acoustic or beam mode waves. 
\end{itemize}

Those waves are observed at all of the analyzed shocks despite having different macroscopic parameters. This indicates the fundamental nature of electrostatic waves for energy dissipation at shocks.

\section*{Open Research Section}
MMS data are available at the MMS Science Data Center; see \url{https://lasp.colorado.edu/mms/sdc/public}. Data analysis was performed using the IRFU-MATLAB analysis package \cite{khotyaintsev_2024_11550091}.

\acknowledgments
We thank the MMS team and instrument PIs for data access and support. This work is supported by The Swedish Research Council Grant No. 2018-05514 and The Swedish National Space Agency grant No. 206/19. AL would like to acknowledge support from STFC grant ST/X001008/1.

\bibliography{main}

%Reference citation instructions and examples:
%
% Please use ONLY \cite and \citeA for reference citations.
% \cite for parenthetical references
% ...as shown in recent studies (Simpson et al., 2019)
% \citeA for in-text citations
% ...Simpson et al. (2019) have shown...
%
%
%...as shown by \citeA{jskilby}.
%...as shown by \citeA{lewin76}, \citeA{carson86}, \citeA{bartoldy02}, and \citeA{rinaldi03}.
%...has been shown \cite{jskilbye}.
%...has been shown \cite{lewin76,carson86,bartoldy02,rinaldi03}.
%... \cite <i.e.>[]{lewin76,carson86,bartoldy02,rinaldi03}.
%...has been shown by \cite <e.g.,>[and others]{lewin76}.
%
% apacite uses < > for prenotes and [ ] for postnotes
% DO NOT use other cite commands (e.g., \citet, \citep, \citeyear, \citealp, etc.).
% \nocite is okay to use to add references from your Supporting Information
%

\end{document}